\documentclass[letterpaper,twocolumn]{jpsj3}
\usepackage{txfonts}

\usepackage{graphicx}
\usepackage{color}

\title{%
Absence of  Magnetic Long Range Order in Ba$_3$ZnRu$_2$O$_9$:
A Spin-Liquid Candidate in the $S=3/2$ Dimer Lattice
}

\author{
Ichiro Terasaki$^1$\thanks{terra@cc.nagoya-u.ac.jp},
Taichi Igarashi$^1$,
Takayuki Nagai$^1$,
Kenji Tanabe$^1$,
Hiroki Taniguchi$^1$,
Taku Matsushita$^1$,
Nobuo Wada$^1$,
Atsushi Takata$^2$,
Takanori Kida$^2$,
Masayuki Hagiwara$^2$,
Kensuke Kobayashi$^3$,
Hajime Sagayama$^3$,
Reiji Kumai$^3$,
Hironori Nakao$^3$, and 
Youichi Murakami$^3$
}

\inst{
$^1$Department of Physics, Nagoya University, Nagoya 464-8602, Japan\\
$^2$Center for Advanced High Magnetic Field Science, 
Graduate School of Science,
Osaka University, 
1-1 Machikaneyama, Toyonaka, Osaka 560-0043, Japan\\
$^3$Condensed Matter Research Center and Photon Factory, 
Institute of Materials Structure Science, 
High Energy Accelerator Research Organization, Tsukuba 305-0801, Japan
}

\date{\today}

\abst{
We have discovered a novel candidate
for a spin liquid state in a  ruthenium oxide
composed of dimers of $S = $ 3/2 spins of Ru$^{5+}$,
Ba$_3$ZnRu$_2$O$_9$.
This compound lacks a
long range order  down to 37 mK, 
which is a temperature 5000-times lower than 
the magnetic interaction scale of around 200 K.
Partial substitution for Zn
can continuously vary
the magnetic ground state  
from an antiferromagnetic order
to a spin-gapped state
through the liquid state.
This indicates that 
the spin-liquid state emerges from a delicate balance
of inter- and intra-dimer interactions,
and the spin state of the dimer plays a vital role.
This unique feature should realize a new type of
quantum magnetism.
}

\begin{document} 

\maketitle

Since Anderson proposed the idea of a quantum spin liquid
as a possible ground state 
for a spin-half ($S=1/2$) antiferromagnetic 
triangular lattice \cite{anderson1973}
with a suppressed long-range magnetic ordering due to geometrical
frustration and quantum fluctuations of interacting spins, 
researchers have sought this state of quantum matter \cite{balents2010}.
A quantum spin liquid should possess a ground 
state consisting of highly entangled singlet-spin pairs
and exotic excited states called spinons \cite{lee2005,lee2006}.
Although several candidates have been reported experimentally, 
none has been confirmed.
Organic candidates consist of  ill-defined localized magnetic moments
where the magnetic exchange interaction is
comparable to the charge gap
\cite{shimizu2003,itou2008,isono2014}.
On the other hand, inorganic candidates suffer from unwanted 
disorder/impurity/nonstoichiometry.
Na$_4$Ir$_3$O$_8$ \cite{okamoto2007}
shows a spin-glass-like transition near 6--7 K \cite{dally2014,shockley2015},
whereas ZnCu$_3$(OH)$_6$Cl$_2$ \cite{shores2005}
and Ba$_3$CuSb$_2$O$_9$ \cite{nakatsuji2012}
include a considerable intermixture of cations.
BaCu$_3$V$_2$O$_8$(OH)$_2$  \cite{okamoto2009}
and 6H-B Ba$_3$NiSb$_2$O$_9$ \cite{cheng2011}
have a substantial low-temperature 
Curie tail due to unwanted impurities.
In the case of BaCu$_3$V$_2$O$_8$(OH)$_2$, 
an inhomogeneous  magnetic order 
has been  detected through NMR measurements
around 9 K, below which the unwanted Curie tail grows rapidly
\cite{quilliam2011,hiroi2013}.

We have discovered the absense of magnetic long range order
in a hexagonal lattice of Ru$^{5+}$ 
dimers in Ba$_3$ZnRu$_2$O$_9$ down to 37 mK, 
where neither Curie tail nor glassy behavior
is detected.
The magnetic specific heat shows no anomaly
below 80 K,
and is found to be linear in temperature below around 5 K.
These thermodynamic measurements suggest
a spin-liquid like ground state in this oxide.
The Ru$^{5+}$ ion acts as a well-localized $S=3/2$ spin, 
and the spin liquid is totally
unprecedented in such a large spin.

Polycrystalline samples of Ba$_3M$Ru$_2$O$_9$ 
($M$ = Zn, Co and Ca) were prepared 
with solid state reaction. 
Stoichiometric mixtures of powdered oxide or carbonate sources 
(BaCO$_3$, RuO$_2$, Co$_3$O$_4$, ZnO, CaCO$_3$) 
were ground in an agate mortar, 
and were pre-sintered in air at 1273 K for 12 h. 
The pre-sintered powders were finely ground, pressed into pellets, 
and sintered in air at 1473 K for 72 h.

The synchrotron x-ray diffraction was taken at BL8A\&8B,
Photon-Factory, KEK, Japan.
The energy of the x-ray was adjusted to be 18 keV, which was carefully
calibrated using a standard powder sample of CeO$_2$.
Powder samples were sealed in a silica-glass capillary of
0.1-mm diameter, and the capillary was rotated by an angle of
30 deg from the sample-stage axis during measurement.
The diffraction patterns were analyzed using the Rietveld refinement
with Rietan-FP code \cite{FP}.
The magnetic susceptibility 
was measured with a commercial susceptometer 
(Quantum Design MPMS) above 2 K, 
and with a home-made probe equipped with a SQUID sensor 
in a dilution refrigerator down to 37 mK 
in various external fields up to 7 mT. 
The signal of the sensor was calibrated with the measured data 
using MPMS from 2 to 4 K.
The magnetization at 1.4 K up to 50 T 
was measured in pulsed fields by an induction method 
at Center for Advanced High Magnetic Field Science in Osaka University.
The specific heat was measured with a commercial measurement system 
(Quantum Design PPMS). 
The ac resistivity was measured with an LCR meter (Agilent E4980A) 
with a frequency of 10 kHz. 

\begin{figure}
 \centering
 \includegraphics[width=8cm]{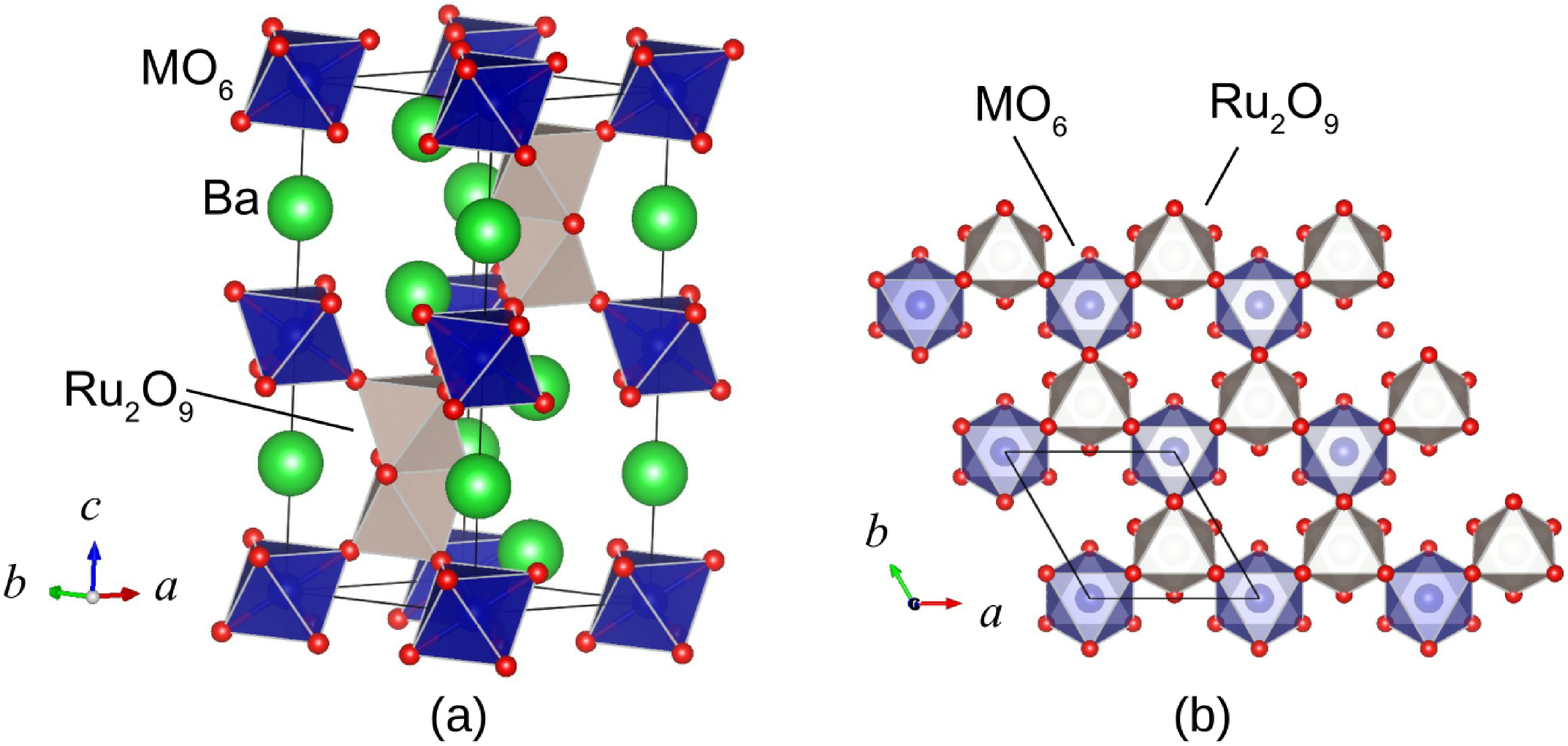}
 \includegraphics[width=6cm]{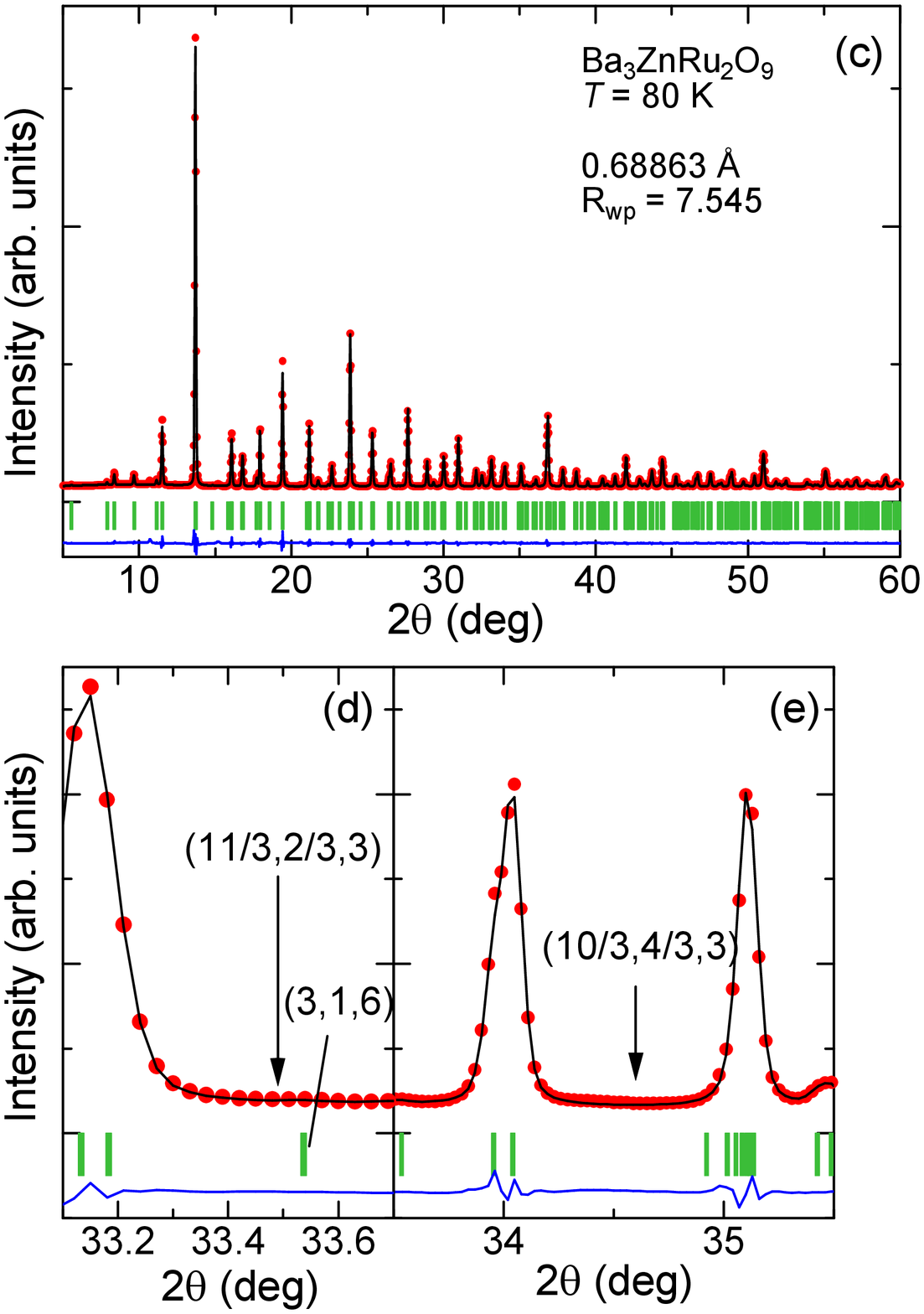}
 \caption{(Color online)
Crystal structure of Ba$_3M$Ru$_2$O$_9$.
(a) Layered structure where the two face-shared RuO$_6$
octahedra (the Ru$_2$O$_9$ dimer) and the $M$O$_6$ octahedron 
are alternately stacked along the $c$-axis.
(b) The $ab$ plane layer where the Ru$_2$O$_9$
dimer and $M$O$_6$ form an edge-shared network
to construct a hexagonal dimer lattice of Ru$^{5+}$.
(c) The synchrotron x-ray diffraction pattern and the Rietveld
 refinement of Ba$_3$ZnRu$_2$O$_9$ at 80 K.
(d)(e) The magnified view of the diffraction pattern.
The arrow indicates a position of possible superlattice reflection
to probe inter-cation mixture of Zn and Ru.
}
\end{figure}

Figure 1(a) schematically shows  the crystal structure of
Ba$_3M$Ru$_2$O$_9$ \cite{lightfoot1990}.
The two face-shared RuO$_6$ octahedra
(Ru$_2$O$_9$ dimer block)
form a layered structure,
and are interconnected through the $M$O$_6$ octahedron 
along the $c$-axis in a corner-shared arrangement.
The Ru$^{5+}$ ion with an electronic
configuration of $(4d)^3$ acts as a local moment of $S=3/2$
and is responsible for the magnetism of Ba$_3M$Ru$_2$O$_9$.

Figure 1(b) depicts the in-layer arrangement of the Ru$_2$O$_9$
dimer blocks,  where each block is connected with three
neighboring $M$O$_6$ octahedra 
to form a hexagonal dimer lattice.
The species of $M$ determines the magnetic ground state.
For $M=$ Co, Ni, and Cu, the system exhibits an antiferromagnetic order
below around $T_N=$ 100 K \cite{lightfoot1990,greenwood1980,cava1999}.
The nearly identical $T_N$ for the different species of $M$
implies that  the magnetic moment of the $M$ ions plays a secondary role.
For $M=$ Ca and Sr, the ground state is a nonmagnetic
spin-gapped state, in which the two localized spins 
in the Ru$_2$O$_9$ dimer block 
form a singlet pair \cite{darriet1976,greenwood1980,attfield2013}.

Figure 1(c) shows the synchrotron x-ray diffraction pattern 
and the Rietveld refinement of Ba$_3$ZnRu$_2$O$_9$ at 80 K.
As shown in the figure, the refinement is satisfactory, and we have
verified the crystal structure shown in Figs. 1(a) and 1(b)
(Space group P63/mmc (No.194),  $a = $ 5.7576 \AA,    $c = $ 14.1424 \AA). 
We also emphasize that no trace of impurity phases is detected, 
and the sample is pure enough to discuss the thermodynamic properties
of the main phase.
Figures 1(d) and 1(e) show the magnified view of the diffraction pattern.
The arrow indicates a position of possible superlattice reflections,
if an inter-mixture of Zn and Ru happened as in the case of
Ba$_3$CuSb$_2$O$_9$ \cite{nakatsuji2012}.
No trace of such reflection peaks safely excludes the possibility of
the Zn-Ru inter-mixture.

\begin{figure}
 \centering
 \includegraphics[width=7cm]{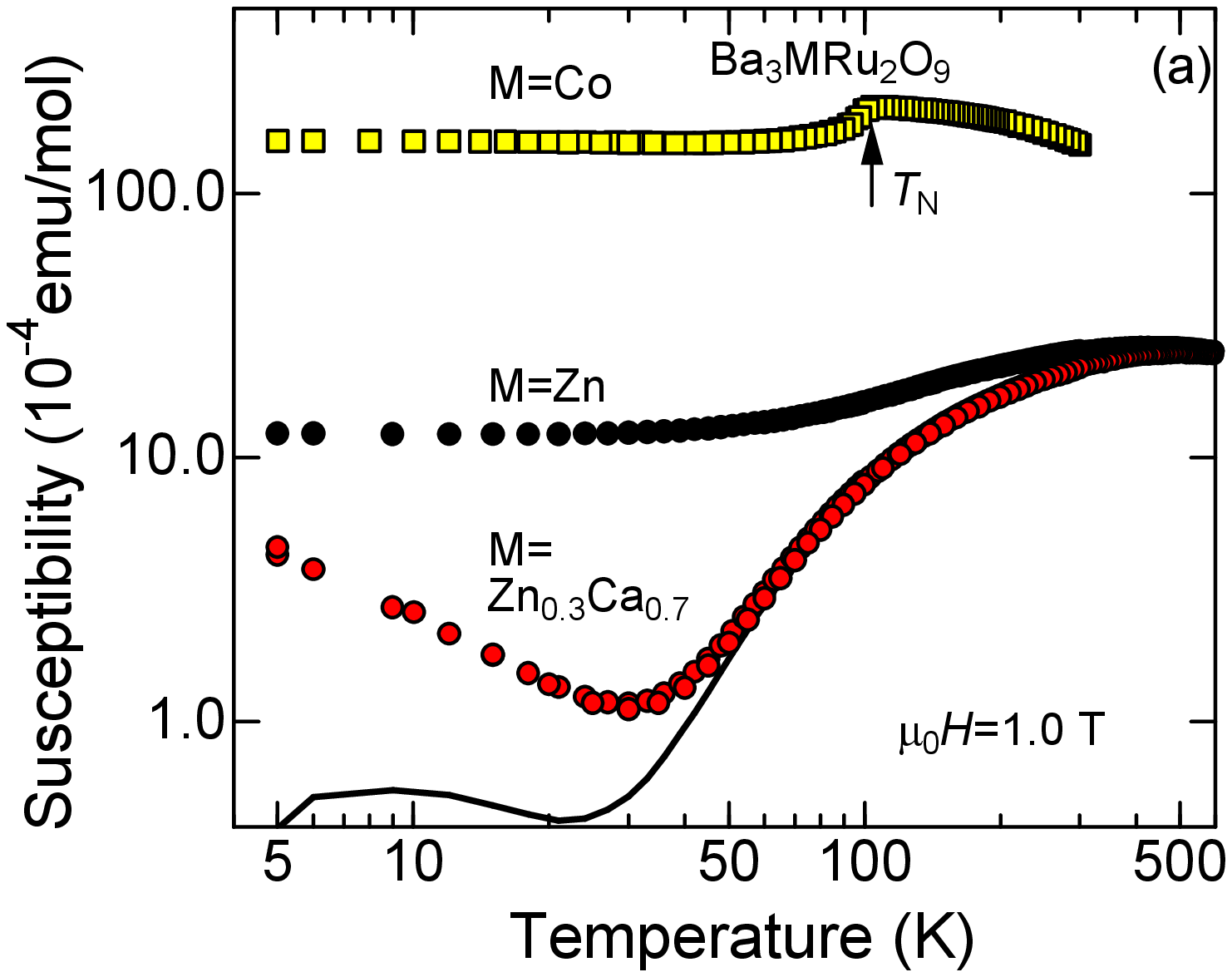}
 \includegraphics[width=5cm]{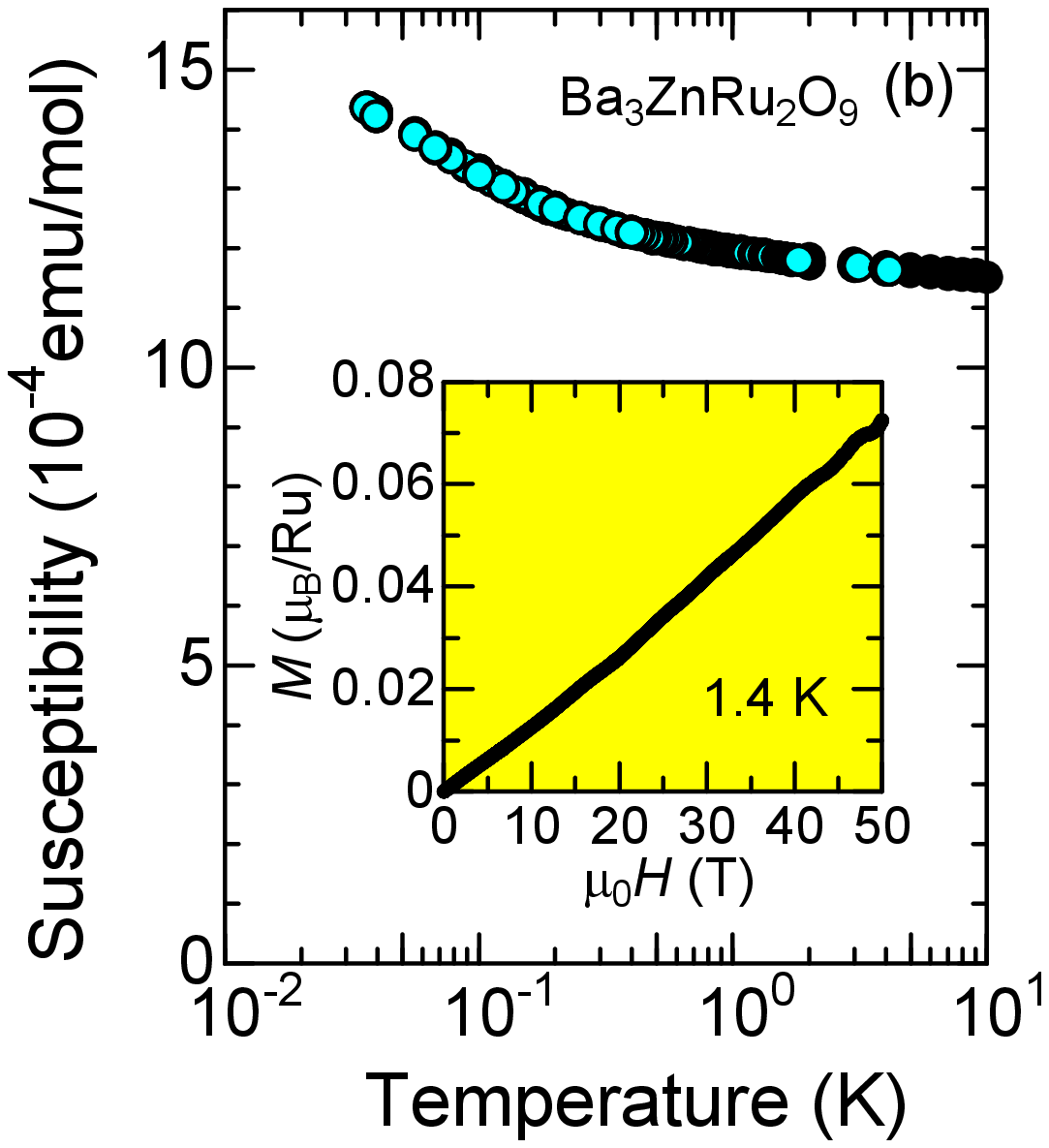}
 \caption{(Color online)
(a) Magnetic susceptibility 
of Ba$_3M$Ru$_2$O$_9$ above 4 K.
Upturn toward low temperatures for
$M=$ Zn$_{0.3}$Ca$_{0.7}$ is due to the 0.1\%
contribution of unwanted magnetic impurities.
Solid curve denotes the intrinsic susceptibility of
$M=$ Zn$_{0.3}$Ca$_{0.7}$ obtained by subtracting 
of the low-temperature Curie term.
(b) Magnetic susceptibility below 10 K for Ba$_3$ZnRu$_2$O$_9$.
No sign of a phase transition is detected down to 37 mK.
The inset shows the magnetization $M$ plotted as a function of
external field $\mu_0H$ at 1.4 K.
}
\end{figure}

Figure 2(a) plots the magnetic susceptibility of Ba$_3M$Ru$_2$O$_9$
for $M =$ Co, Zn, and Zn$_{0.3}$Ca$_{0.7}$  on a logarithmic scale.
Although  an antiferromagnetic transition occurs
 for the $M=$ Co sample as a kink around 100 K,
a much smaller susceptibility is observed for
the $M=$ Zn$_{0.3}$Ca$_{0.7}$ sample.
If  the low-temperature Curie tail is subtracted 
by assuming a tiny contribution (0.1 mol \%)
of an unwanted magnetic impurity, 
the solid curve shows the intrinsic susceptibility.
The curve gives a nearly temperature-independent value
of 10$^{-5}$ emu/mol below 20 K,
which can be assigned to the Van Vleck susceptibility
of the Ru ions.

The $M=$ Zn sample  shows a 
magnetic susceptibility intermediate between 
the $M=$ Co and Zn$_{0.3}$Ca$_{0.7}$ samples,
which is roughly consistent with a very recent report 
\cite{beran2015}.
This sample shows a weakly temperature-dependent susceptibility
with a broad maximum around 400 K
without  any trace of a cusp or discontinuity down to 37 mK
(Fig. 2(b)).
A Curie tail is not visible throughout the measured temperature ranges, 
whereas the susceptibility weakly increases with decreasing temperature
below 1 K.  
The inset of Fig. 2(b) shows the magnetization plotted as a function
of external field up to 50 T at 1.4 K.
The linear and reversible behaviour indicates
an antiferromagentic interaction larger than 50 T. 
If magnetic impurities were present, 
the magnetization curve should be convex upward 
as can be explained with the Brillouin function.
If the system were in a glassy state
as seen in BaCu$_3$V$_2$O$_8$(OH)$_2$,\cite{okamoto2009} 
the nonlinear susceptibility (i.e. the magnetization devided by external field)
should decrease with increasing external field.
The present magnetization curve excludes a possible existence of 
free magnetic impurities and/or glassy states.
Previous neutron diffraction studies have also reported
no long range order \cite{lightfoot1990,beran2015}.
Therefore, these results strongly indicate that this paramagnetism 
is due to a quantum spin-liquid state.

Let us evaluate the magnitude of the intra-dimer exchange energy.
Darret et al. \cite{darriet1976} first analyzed 
the susceptibility of Ba$_3M$R$_2$O$_9$ ($M$ = Ca, Mg, Cd)
assuming  independent spin dimers, and evaluated 
the intra-dimer interaction $J_{\rm intra}$ to be around 200 K. 
Later Senn et al. evaluated $J_{\rm intra}$ to be 240 K 
for Ba$_3$CaRu$_2$O$_9$ \cite{attfield2013}. 
In an LDA+U calculation for Ba$_3$CoRu$_2$O$_9$, 
Streltsov \cite{streltsov2013} 
theoretically evaluated  $J_{\rm intra}$ to be 150 K.
Thus we can conclude that magnetic phase transition is not 
detected at temperatures 5000 times lower than the interaction energy
scale in Ba$_3$ZnRu$_2$O$_9$.

\begin{figure}
 \centering
 \includegraphics[width=8cm]{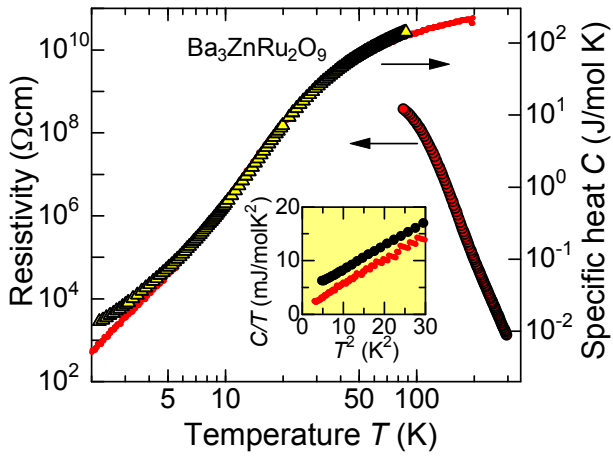}
 \caption{(Color online)
Resistivity and specific heat of
Ba$_3$ZnRu$_2$O$_9$ plotted as a function of temperature $T$.
Inset plots $C/T$ as a function of $T^2$
where the gapless $T-$linear contribution of $C$ is
evaluated to be 4 mJ/mol K$^2$.
The small dotted data represent the specific heat of 
a non-magnetic reference material Ba$_3$ZnSb$_2$O$_9$
taken from Ref. \cite{cheng2011}.
}
\end{figure}

Figure 3 shows the resistivity and specific heat
of Ba$_3$ZnRu$_2$O$_9$.
The resistivity is as high as 10$^3$ $\Omega$cm at room temperature,
and increases up to 10$^{8}$ $\Omega$cm at 100 K.
This highly insulating behavior eliminates the possibility that
the susceptibility of this oxide is due to simple
Pauli paramagnetism.
The temperature dependence near 300 K 
implies  that the activation energy is around 2000 K, 
which greatly exceeds $J_{\rm intra} = $ 150--240 K.
These observations indicate that the electrons 
are well localized in this oxide, justifying that localized magnetic moments 
are responsible for the magnetism.

The specific heat does not  show an anomaly from 80 down to 2 K, 
indicating the absence of a thermodynamic phase transition.
The inset shows  the existence of a temperature-linear contribution 
in the specific heat, which is evaluated to be 4 mJ/mol K$^2$
for $T\to$ 0.
The specific heat does not exhibit a Schottky anomaly above 2 K,
which is consistent with the absence of
a Curie tail in the susceptibility.
The linear dependence is most likely from spinons and not
simple magnons.
The magnitude is on the same order as the electron specific heat
coefficient of alkaline metals, providing evidence of
gapless excitations in the spin sector.
In order to examine the magnetic contribution, we plot the specific
heat of Ba$_3$ZnSb$_2$O$_9$ taken from Ref. \cite{cheng2011}
as a non-magnetic reference material.
As is clearly seen, the specific heat of the two samples 
is almost identical above 10 K, and the magnetic contribution 
of Ru is hidden in the overwhelming major contribution from the lattice. 
The inset shows that the specific heat of Ba$_3$ZnSb$_2$O$_9$
has the identical $T^3$ term without $T$-linear term.
This indicates that the magnetic contribution is proportional to $T$
in Ba$_3$ZnRu$_2$O$_9$ at low temperatures.

\begin{figure}
 \centering
 \includegraphics[width=7cm]{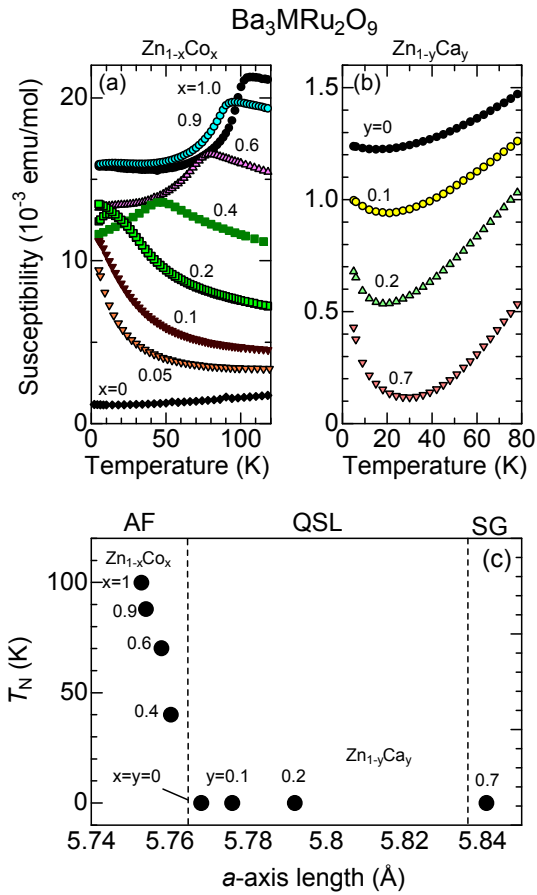}
 \caption{(Color online)
Magnetic susceptibility of 
(a) Ba$_3$Zn$_{1-x}$Co$_x$Ru$_2$O$_9$ and 
(b) Ba$_3$Zn$_{1-y}$Ca$_y$Ru$_2$O$_9$.
(c) Transition temperature plotted as 
a function of the $a$-axis length (the inter-dimer distance).
Dotted lines indicate the approximate
phase boundaries determined from (a) and (b).
AF, QSL, and SG stand for antiferromagnetically ordered state,
quantum spin-liquid state and spin-gapped state,
respectively.
}
\end{figure}

A unique feature of Ba$_3M$Ru$_2$O$_9$ is that the magnetic
ground state can be finely tuned from the antiferromagnetic
order to the non-magnetic spin-gapped state through
the gapless quantum spin-liquid state.
Figures 4(a) and (b) show the magnetic susceptibility
of Ba$_3$Zn$_{1-x}$Co$_x$Ru$_2$O$_9$ and 
Ba$_3$Zn$_{1-y}$Ca$_y$Ru$_2$O$_9$,
respectively.
The susceptibility at 120 K systematically increases with 
the Co concentration $x$, because the magnetic Co$^{2+}$
ion contributes to the susceptibility in Ba$_3$Zn$_{1-x}$Co$_x$Ru$_2$O$_9$.
Simultaneously, the antiferromagnetic transition temperature
systematically increases with $x$.
The 20\% Co substituted sample ($x = 0.2$)
shows a temperature hysteresis below approximately 15 K,
which can be assigned to a glass transition.
For $x < 0.2$, the system seems to be in a spin-liquid state.
For Ca substitution, the $y = 0.2$ sample 
shows a finite paramagnetic contribution around 20 K
and is  still in the spin-liquid state.
The magnitude continuously decreases with $y$,
implying a magnetically inhomogeneous state 
such as a mixture of spin-liquid and spin-gapped states.

Here we discuss a possible microscopic mechanism to cause various
magnetic ground states in Ba$_3M$Ru$_2$O$_9$.
Figure 4 (c) plots the transition temperature as
a function of the $a$-axis length, or equivalently, 
the inter-dimer distance. 
The two dotted lines represent the phase boundary
determined from Figs. 4(a) and 4(b).
First of all, we emphasize that the inter-dimer interaction 
competes with the intra-dimer interaction ($J_{\rm intra}$=150--240 K) 
in this family,
if the number of the nearest neighbor dimers ($z$ = 6) 
is taken into account.
According to the LDA+U calculation,\cite{streltsov2013}
the interaction $J_{\rm inter}$ along the Ru--O--O--Ru network
between the neighboring dimers
is evaluated to be 30--40 K, and we arrive at 
the particular condition $zJ_{\rm inter} \sim J_{\rm intra}$.
We further note that the interaction between Co and Ru is 
evaluated to be less than 8 K, being much smaller than $J_{\rm inter}$.
Thus, even though the magnetism of Co$^{2+}$ ions complicates
the magnetic structure, we may safely ignore this in the lowest
approximation.
We expect that the $a$-axis length changes $J_{\rm inter}$;
For a short $a$, the condition $zJ_{\rm inter} > J_{\rm intra}$ 
stabilizes the antiferromagnetic order, wheres for a long $a$,
$zJ_{\rm inter} < J_{\rm intra}$ favors the spin singlet within
dimers.
We expect that the quantum spin-liquid state can emerge from 
a delicate condition of $zJ_{\rm inter} \sim J_{\rm intra}$.

In conventional dimer spin lattices,  the spin-liquid state is 
expected to emerge just at the boundary
between the spin-gapped state and the antiferromagnetic state.
In the present system, in contrast, the spin-liquid state seems to 
be stable in a finite range of $zJ_{\rm inter}/J_{\rm intra}$.
We think that this arises from the fact that 
the magnetic moment is comprised of
dimerized spins; the two $S=3/2$ spins can take spin states of 
$S_{\rm tot}=$  0, 1, 2, or 3.
An early neutron experiment \cite{lightfoot1990}
and first-principles calculations \cite{streltsov2013}
suggest $S_{\rm tot}\sim$ 2 (1.5-2$\mu_B$ per Ru)
for $M$ = Co, but $S_{\rm tot}=$ 0
for $M$ = Ca \cite{attfield2013}. 
Hence, we expect that the magnitude of $S_{\rm tot}$ 
varies (perhaps dynamically) between 0 and 2
for $M$ = Zn, causing a strong magnetic fluctuation, 
which suppresses the magnetic ordering.
A recent theoretical study by Watanabe et al. \cite{watanabe2014}
suggests that randomness in the exchange interaction
can induce spin-liquid-like  behavior.
The present oxide may have randomness not 
in $J_{\rm inter}$ but in $S_{\rm tot}$.   
This idea can be examined by carefully analyzing the neutron
diffraction of Ba$_3$Zn$_{1-x}$Co$_x$Ru$_2$O$_9$.

In summary, we have discovered that no magnetic transitions
occur down to 37 mK in Ba$_3$ZnRu$_2$O$_9$.
The $T$-linear magnetic specific heat and the paramagnetic
susceptibility strongly suggest a quantum spin liquid state.
Considering that the related oxides show the antiferromagnetic order
or the spin-gapped nonmagnetic state,
we suggest that competing interaction between
intra- and inter-dimer interactions should stabilize
this spin liquid like state.

The authors would like to thank Chisa Hotta and Yukio Yasui
 for fruitful discussion and useful advise.
This work was partially supported by 
Grant-in-Aid for Scientific Research and a Grant-in-Aid for JSPS Fellows, 
Japan Society for the Promotion of Science, Japan 
(Kakenhi Nos. 25610091, 26247060, 15J04615),
and by Program for Leading Graduate Schools ``Integrative Graduate
Education and Research in Green Natural Sciences'', MEXT, Japan.
The synchrotron x-ray diffraction was performed under
the approval of the Photon Factory Program
Advisory Committee (Proposal Nos. 2012G718 and 2012S2-005).


\end{document}